\documentclass[11pt]{asaproc}
\pdfoutput=1
\usepackage{times}
\usepackage{amsmath,amsfonts}
\usepackage{url,verbatim,listings,graphicx,listings}
\usepackage{natbib}

\def\eatzero#1{\ifnum#1=0\relax\else{#1}\fi}
\def\eattwo#1#2{\relax}
\def\svndate$#1: #2-#3-#4 #5 #6 (#7) ${\def\fmtdate{#2--#3--#4}}
\svndate$Date: 2010-11-16 14:11:46 -0500 (Tue, 16 Nov 2010) $

\DeclareMathOperator{\Uniform}{Uniform}
\newcommand{\R}{\mathbb{R}}

\title{Slice Sampling with Adaptive Multivariate Steps:\\
The Shrinking-Rank Method}
\author{Madeleine B. Thompson\thanks{Department of Statistics,
University of Toronto} \and Radford M. Neal \thanks{Department of
Statistics and Department of Computer Science, University of Toronto}}
\date{\fmtdate}

\begin{document}

\maketitle

\begin{abstract}
The shrinking rank method is a variation of slice sampling that is
efficient at sampling from multivariate distributions with highly
correlated parameters.  It requires that the gradient of the
log-density be computable. At each individual step, it approximates
the current slice with a Gaussian occupying a shrinking-dimension
subspace.  The dimension of the approximation is shrunk orthogonally
to the gradient at rejected proposals, since the gradients at points
outside the current slice tend to point towards the slice.  This
causes the proposal distribution to converge rapidly to an estimate
of the longest axis of the slice, resulting in states that are less
correlated than those generated by related methods.  After describing
the method, we compare it to two other methods on several distributions
and obtain favorable results.
\end{abstract}

\section{Introduction}

Many Markov Chain Monte Carlo methods mix slowly when parameters
of the target distribution are highly correlated; many others mix
slowly when the parameters have different scaling.  This paper
describes a variation of slice sampling \citep{neal03}, the
\textit{shrinking-rank method}, that performs well in such
circumstances.  It assumes the parameter space is continuous and
that the log-density of the target distribution and its gradient
are computable.  We will first describe how the method works, then
compare its performance, robustness, and scalability to two other
MCMC methods.

\section{Description of the shrinking-rank method}
\label{shrinking-rank}

\begin{figure}
\begin{center}\begin{tabular}{cc}\includegraphics{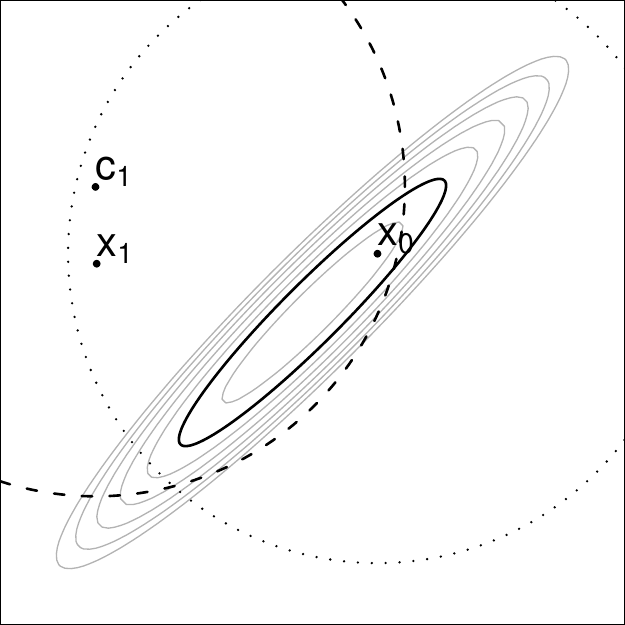} &
  \includegraphics{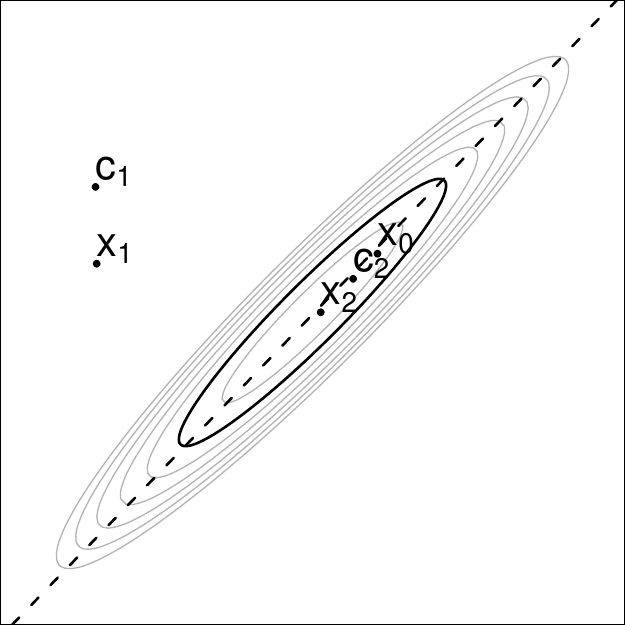} \\ (a) & (b)\end{tabular}\end{center}
\caption{(a) The grey lines represent the contours of a two-dimensional
distribution; the solid ellipse represents the boundary of the
slice.  The first crumb, $c_1$, is drawn from a spherical Gaussian
represented by a dotted circle; a proposal, $x_1$, is drawn from a
spherical Gaussian centered at $c_1$, represented by a dashed circle.
$x_1$ is rejected because it is outside the solid ellipse.  (b) A
second crumb, $c_2$, is drawn from a reduced-rank subspace, represented
by a dashed line.  A second proposal, $x_2$, is drawn from the same
subspace.  Since $x_2$ is inside the solid ellipse, it is accepted.}
\label{crumb-proposal-figure}
\end{figure}

Suppose we wish to sample from a target distribution with density
function $f(\cdot)$, and the current state is $x_0$.  In slice
sampling, we first draw a slice level, denoted by $y$, uniformly
from the interval $[0,f(x_0)]$.  Then, we update $x_0$ in a
way that leaves the uniform distribution on the slice $\{ x | f(x)
\geq y\}$ invariant.  The resulting stationary distribution of the
$(x,y)$ pairs is uniform on the area underneath $f(\cdot)$, and the
marginal distribution of the $x$ coordinates has density $f(\cdot)$,
as desired.

The crumb framework of slice sampling \citep[\S5.2]{neal03} is a
particular way of updating $x_0$.\footnote{In the interest of
brevity, we have omitted a full description of the crumb framework.
Readers interested in understanding the correctness of the method
described in this paper may find \citet[\S5.2]{neal03} and
\citet{thompson10} helpful.}  First, we draw a crumb from some
distribution (to be specified later).  Then, we propose a new state
from the distribution of states that could have generated that
crumb.  If the proposal is in the slice, we accept the proposal as
the new state.  Otherwise, we draw further crumbs and proposals
until a proposal is in the slice.

In the shrinking-rank method, the crumbs are Gaussian random variables
centered at the current state.  To ensure that the uniform distribution
on the slice is invariant under state transitions, we will make the
probability of starting at $x_0$ and accepting a proposal $x_k$ the
same as the probability of starting at $x_k$ and accepting $x_0$.
This requirement is satisfied if proposal $k$ is drawn from a
Gaussian with precision equal to the sum of the precisions of
crumbs $1$ to $k$ and mean equal to the precision-weighted mean
of crumbs $1$ to $k$.

Further, the precision matrices may depend arbitrarily on the
locations and densities of the previous proposals; we take advantage of this
by choosing crumb precision matrices that result in state transitions
that take large steps along the slice.  When the first crumb, $c_1$,
is drawn, there are no previous proposals providing information
to adapt on, so we draw it from a spherical Gaussian distribution with
standard deviation $\sigma_c$, where $\sigma_c$ is a tuning parameter.
The distribution for the first proposal, $x_1$, is also a spherical
Gaussian with standard deviation $\sigma_c$, but centered at $c_1$
instead of $x_0$.

If $x_1$ is outside the slice, we can use the gradient of the log
density at $x_1$ to determine a distribution for $c_2$ that leads
to a distribution for $x_2$ that more closely resembles the shape
of the slice itself.  In particular, we consider
setting the variance of the distribution of $c_2$ to be zero in the
direction of the gradient, since the gradients are orthogonal to
the contours of the log density.  If the contour defined by the log
density at the proposal and the contour defined by the the slice
level are the same shape, this will result in a crumb, and therefore
a proposal, being drawn from a distribution oriented along the long
directions of the slice.  This procedure is illustrated in
figure~\ref{crumb-proposal-figure}.

The nullspace of the subspace the next crumb is to be drawn from
is represented by $J$, a matrix with orthogonal, unit-length columns.
Let $g^*$ be the projection of the gradient of the log density at
a rejected proposal into the nullspace of $J$.  When $g^*$ makes a
large angle with the gradient, it does not make sense to adapt based
on it, because this subspace is already nearly orthogonal
to the gradient.  When the angle is small, we extend $J$ by appending
$g^*/\lVert g^*\rVert$ to it as a new column.  Here, we define a
large angle to be any angle greater than $60^\circ\!$, but the
exact value is not crucial.

Formally, define $P(J,v)$ to be the projection of vector $v$ into
the nullspace of the columns of $J$ (so that it returns vectors in
the space that crumbs and proposals are drawn from):
\begin{equation}\label{P}
P(J,v) = \begin{cases} v - J J^T v & \quad \text{if $J$ has at
least one column} \\ v & \quad \text{if $J$ has no columns} \end{cases}
\end{equation}
We let $g^*$ be the projection of the gradient at the proposal orthogonal
to the columns of $J$:
\begin{equation*}
g^* = P\left(J, \nabla \log f(x_k) \right)
\end{equation*}
Then we update $J$ if
\begin{equation*}
\frac{{g^*}^T \nabla\log f(x_k)}{\lVert g^* \rVert \,
  \lVert \nabla\log f(x_k) \rVert} > \cos 60^\circ
\end{equation*}
and the nullspace of $J$ is not one dimensional.  This update to $J$ is:
\begin{equation*}
J \leftarrow \left[ J \quad \frac{g^*}{\lVert g^* \rVert} \right]
\end{equation*}

To ensure a proposal is accepted in a reasonable number of iterations,
if we do not update $J$ for a particular crumb, we scale down
$\sigma_c$ by a configurable parameter $\theta$ (commonly set to
0.95).  Write the standard deviation for the $k$th crumb as
$\sigma_{c(k)}$.  If we never updated $J$, then $\sigma_{c(k)}$
would equal $\theta^{k-1} \sigma_c$.  Since we only change one of
$J$ or the standard deviation each step, $\sigma_{c(k)}$ does not
fall this fast.  If the standard deviation were updated every step,
it would fall too fast in high-dimensional spaces where many updates
to $J$ are required before the proposal distribution is reasonable.
As a further refinement, we down-scale $\sigma_{c(k)}$ by an
additional factor of $0.1$ when the density at a proposal is
zero.  Since the usual form of adaptation is not possible in
this case, this scaling results in significantly fewer crumbs and
proposals on distributions with bounded support.

After drawing the $k$th crumb the mean of the distribution for the
next proposal is:
\begin{equation*}
x_0 + P\left(J, \; \frac{\sigma^{-2}_{c(1)} (c_1-x_0) +
\cdots + \sigma^{-2}_{c(k)} (c_k-x_0)}{\sigma^{-2}_{c(1)} + \cdots
+ \sigma^{-2}_{c(k)}} \right)
\end{equation*}
The mean of the proposal distribution is computed as an offset to
$x_0$, but any point in the nullspace of the columns of $J$ would
generate the same result.  In that space, the offset of the proposal
mean is the mean of the offsets of the crumbs weighted by their
precisions.  The variance of the proposals in that space is the
inverse of the sum of the precisions of the crumbs:
\begin{equation*}
\left( \sigma^{-2}_{c(1)} + \cdots + \sigma^{-2}_{c(k)} \right)^{-1}
\end{equation*}

One shrinking rank slice sampler update is shown in
figure~\ref{pseudocode}.  This will be repeated every iteration of
the Markov chain sampler.  It could be combined with other updates,
but we do not consider this here.

\begin{figure}
\begin{lstlisting}[mathescape,frame=single,columns=fullflexible,
title=\textbf{One step in the shrinking-rank method},escapechar=|,lineskip=2pt]
$y \leftarrow \Uniform(0, f(x_0))$
$k \leftarrow 0$
$\sigma_{c(1)} \leftarrow \sigma_c$
$J \leftarrow [ \;\; ]$
repeat until a proposal is accepted:
  $k \leftarrow k + 1$
  $c_k \leftarrow P\bigl(J, N(x_0,\sigma_{c(k)}^2 I)\bigr)$
  $\sigma_x^2 \leftarrow
    \left( \sigma^{-2}_{c(1)} + \cdots + \sigma^{-2}_{c(k)} \right)^{-1}$
  $\mu_x \leftarrow \sigma_x^2 \left( \sigma^{-2}_{c(1)} (c_1-x_0) + \cdots +
    \sigma^{-2}_{c(k)} (c_k-x_0) \right)$
  $x_k \leftarrow x_0 + P\left(J, N(\mu_x, \sigma_x I) \right)$
  if $f(x_k) \geq y$:
    accept proposal $x_k$
  end (if)
  $g^* \leftarrow P(J, \nabla \log f(x_k))$
  if $J$ has fewer than $p-1$ columns and ${g^*}^T \nabla\log f(x) >
      \cos(60^\circ) \cdot \lVert g^*\rVert \,
      \lVert \nabla \log f(x) \rVert$:
    $J \leftarrow [ \; J \quad g^*/\lVert g^*\rVert \; ]$
    $\sigma_{c(k+1)} \leftarrow \sigma_{c(k)}$
  else
    $\sigma_{c(k+1)} \leftarrow \theta \cdot \sigma_{c(k)}$
  end (if)
end (repeat)
\end{lstlisting}
\caption{This pseudocode represents a single transition in the
shrinking rank method with density function $f$, starting from state
$x_0 \in \R^p$, and with tuning parameters $\sigma_c$ and $\theta$.
The mean and variance of the proposals inside the nullspace of $J$
are $\mu_x$ and $\sigma_x^2$.  The density of the current
slice level is $y$; a real implementation would use the log density.
The projection function, $P$, is defined in equation~\ref{P}.
The function $N(\mu,\Sigma)$ generates a multivariate Gaussian with
mean $\mu$ and covariance $\Sigma$.}\label{pseudocode}
\end{figure}

\section{Comparison with other methods}
\label{comparison-section}

\begin{figure}
\begin{center}\includegraphics[height=5in]{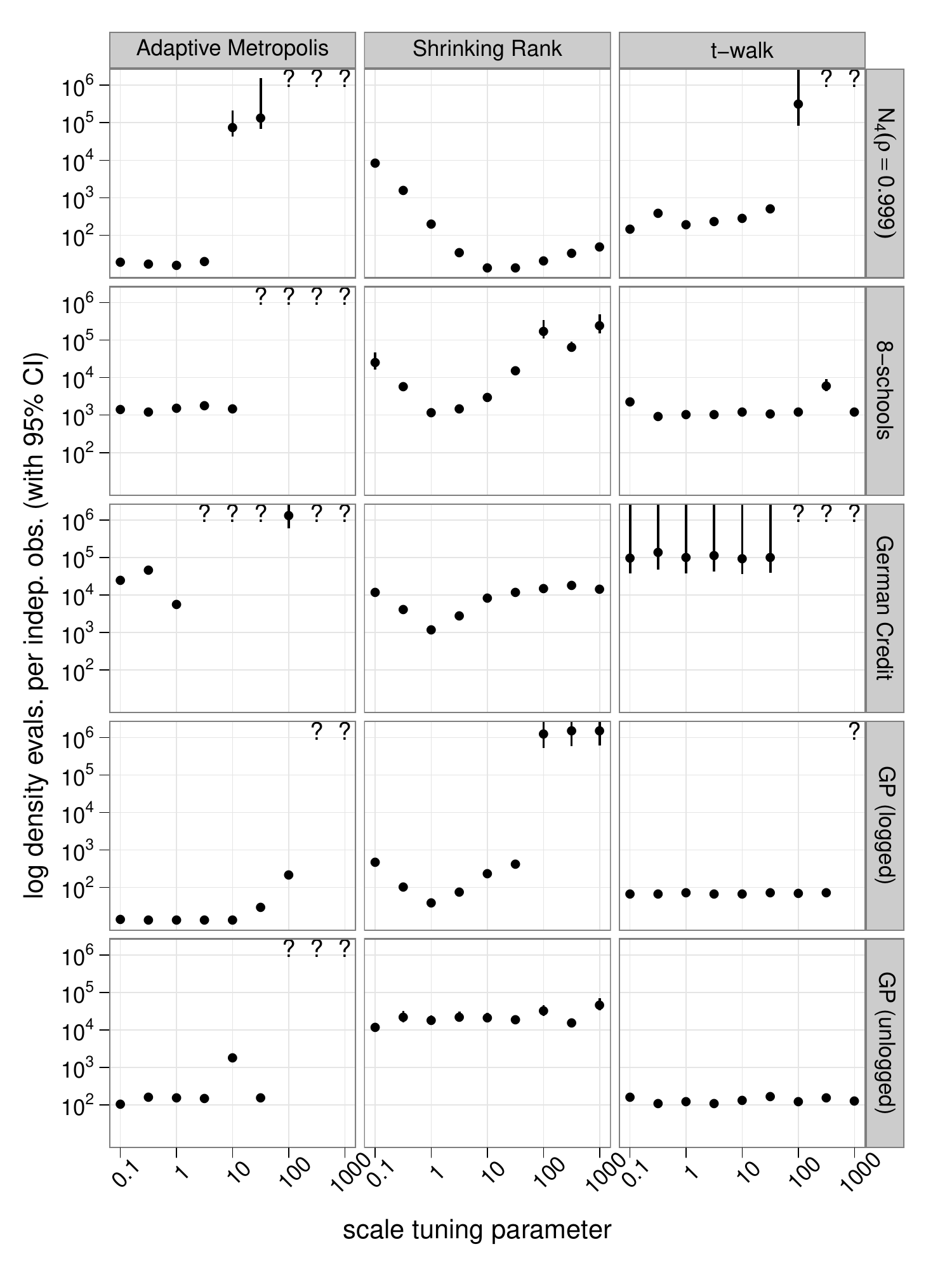}\end{center}
\caption{A comparison of three samplers on five distributions using
simulations of length 200,000.  Log density evaluations per independent
observation (lower is better) are plotted against each distribution's
tuning parameter,
with asymptotic 95\% confidence intervals shown as bars (sometimes
too short to be visible).  Question marks indicate simulations that
had fewer than five distinct observations---too few for the
autocorrelation time to be estimated.  See section~\ref{comparison-section}
for a description of the distributions and samplers.  See
\citet{corlen} for discussion of this type of plot.}
\label{comparison-figure}
\end{figure}

\begin{figure}
\begin{center}\includegraphics[height=3.2in]{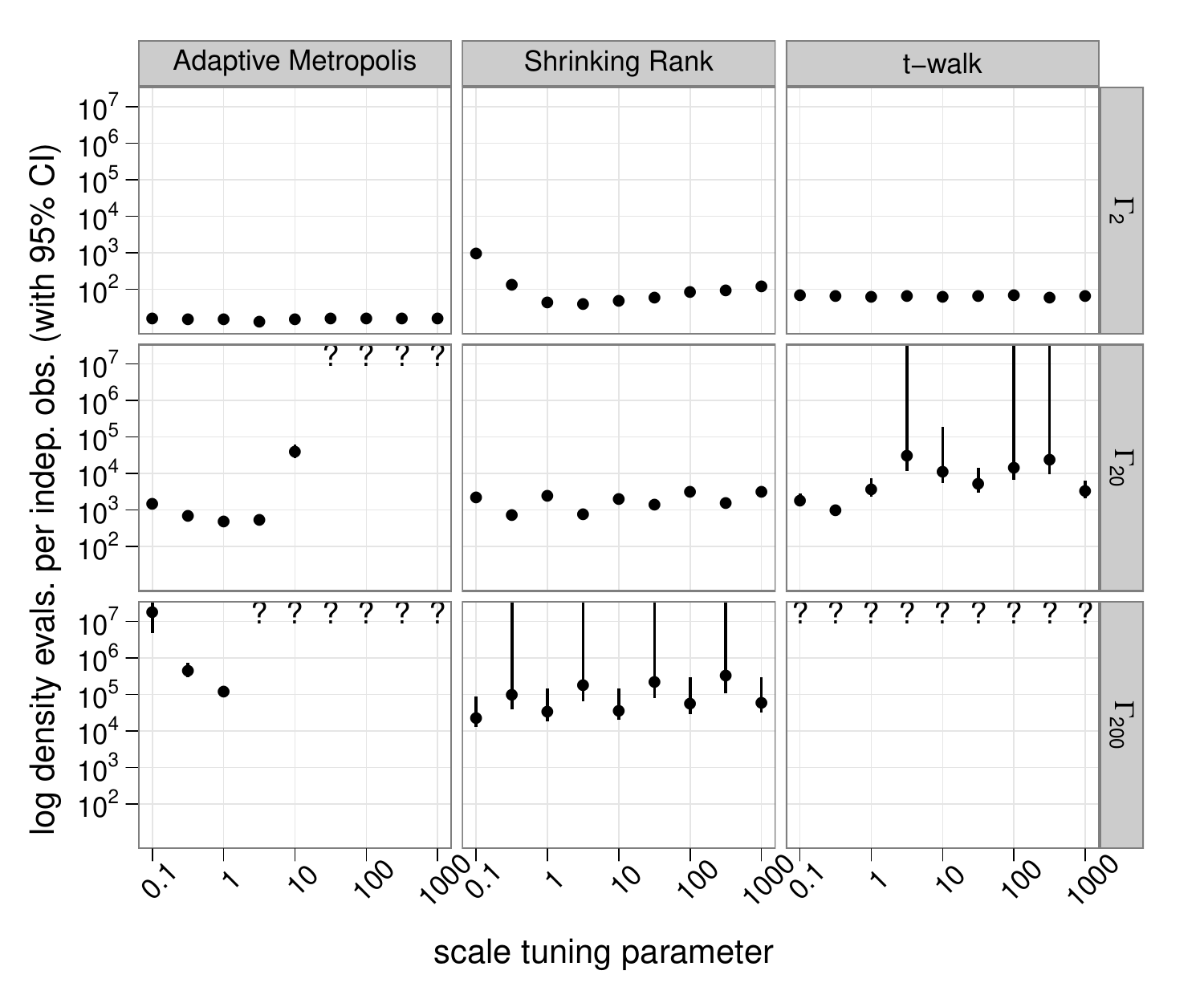}\end{center}
\caption{A comparison of three samplers on distributions with
uncorrelated Gamma(2,1) marginals.  The three distributions have
dimensions 2, 20, and 200.  Each simulation is of length 60,000.}
\label{scaling-figure}
\end{figure}

Figure~\ref{comparison-figure} compares the shrinking-rank method
to two other MCMC methods: t-walk and Adaptive Metropolis.  The
t-walk, described in \citet{christen10}, has a tuning parameter
that specifies the separation of the initial coordinate pair.
Adaptive Metropolis \citep{roberts09} takes multivariate steps with
a proposal covariance matrix chosen based on previous states.  Its
tuning parameter is the standard deviation of its initial proposal
distribution multiplied by the square root of the problem dimension.
The shrinking-rank method is described in section~\ref{shrinking-rank}.
The tuning parameter that is varied is $\sigma_c$; $\theta$ is fixed
at $0.95$.

\newpage

We compare these methods using five distributions:
\begin{itemize}
\item $N_4(\rho=0.999)$: a four dimensional Gaussian with
  highly-correlated parameters; the covariance matrix has condition
  number 2800.
\item Eight Schools \citep[pp.~138--145]{gelman04}: a well-conditioned
  hierarchical model with ten parameters.
\item German Credit \citep[p.~15]{girolami09}: a Bayesian logistic
  regression with twenty-five parameters.  The data matrix is not
  standardized.
\item GP (logged) and GP (unlogged): a Bayesian Gaussian process
  regression with three parameters: two variance components and a
  correlation decay rate.  Its contours are not axis-aligned.  The
  unlogged variant is right skewed in all parameters; the logged
  variant, in which all three parameters are log-transformed, is more
  symmetric.
\end{itemize}

The shrinking rank method tends to perform well for a wide range
of tuning parameters on the first three distributions.  Adaptive
Metropolis also performs well, as long as the tuning parameter is
smaller than the square root of the smallest eigenvalue of the
target distribution's covariance.  The recommended value, 0.1, would
have worked well for all three distributions.  The t-walk works
well on the low dimensional distributions, but fails on the
higher-dimensional German credit distribution.

The inferior performance of the shrinking rank method on the unlogged
Gaussian process regression shows one of its weaknesses: it does not
work well on highly skewed distributions because the gradients at
rejected proposals often do not point towards the slice.  As can
be seen by comparing to the logged variation, removing the skewness
improves its performance substantially.

Figure~\ref{scaling-figure} shows a set of simulations on distributions
of increasing dimension, where each component is independently
distributed as Gamma(2,1).  For the shrinking rank method and
Adaptive Metropolis, multiplying the dimension by ten corresponds
roughly to a factor of ten more function evaluations.  The t-walk
does not scale as well.  A similar experiment using standard Gaussians
instead of Gamma distributions gives equivalent results.

\section{Discussion}

The main disadvantage of the shrinking rank method is that it can only
be used when the gradient of the log density is available.  One
advantage is that it is rotation and translation invariant, and
nearly scale invariant.  It performs comparably to Adaptive Metropolis,
but unlike Adaptive Metropolis, adapts to local structure each
iteration instead of constructing a single proposal distribution.

An R implementation of the shrinking rank method and the Gaussian
process distribution from section~\ref{comparison-section} can be
found at \url{http://www.utstat.toronto.edu/mthompson}.
A C implementation of the shrinking rank method will be included in the
forthcoming SamplerCompare R package.  The shrinking rank method
and a related method, covariance matching, are also discussed in
\citet{thompson10}.

\bibliographystyle{apalike}
\bibliography{../corlen/corlen}

\begin{thebibliography}{}

\bibitem[Christen and Fox, 2010]{christen10}
Christen, J.~A. and Fox, C. (2010).
\newblock A general purpose sampling algorithm for continuous distributions
  (the t-walk).
\newblock {\em Bayesian Analysis}, 5(2):1--20.

\bibitem[Gelman et~al., 2004]{gelman04}
Gelman, A., Carlin, J.~B., Stern, H.~S., and Rubin, D.~B. (2004).
\newblock {\em Bayesian Data Analysis, Second Edition}.
\newblock Chapman and Hall/CRC.

\bibitem[Girolami and Calderhead, 2011]{girolami09}
Girolami, M. and Calderhead, B. (2011).
\newblock {Riemann} manifold {Langevin} and {Hamiltonian} {Monte} {Carlo}.
\newblock {\em Journal of the Royal Statistical Society B}, 73:1--37.
\newblock arXiv:0907.1100v1 [stat.CO].

\bibitem[Neal, 2003]{neal03}
Neal, R.~M. (2003).
\newblock Slice sampling.
\newblock {\em Annals of Statistics}, 31:705--767.

\bibitem[Roberts and Rosenthal, 2009]{roberts09}
Roberts, G.~O. and Rosenthal, J.~S. (2009).
\newblock Examples of adaptive {MCMC}.
\newblock {\em Journal of Computational and Graphical Statistics},
  18(2):349--367.

\bibitem[Thompson, 2010]{corlen}
Thompson, M.~B. (2010).
\newblock Graphical comparison of {MCMC} performance.
\newblock Technical Report 1010, Dept. of Statistics, University of Toronto.
\newblock arXiv:1011.4457v1 [stat.CO].

\bibitem[Thompson and Neal, 2010]{thompson10}
Thompson, M.~B. and Neal, R.~M. (2010).
\newblock Covariance-adaptive slice sampling.
\newblock Technical Report 1002, Dept. of Statistics, University of Toronto.
\newblock arXiv:1003.3201v1 [stat.CO].

\end{thebibliography}

\end{document}